\documentclass[pdfa,a4paper,USenglish,cleveref,autoref,thm-restate]{lipics-v2021}

\pdfoutput=1 %uncomment to ensure pdflatex processing (mandatatory e.g. to submit to arXiv)
\title{Amortized Analysis via Coinduction}
\category{Early Ideas}

\author{Harrison Grodin}{Computer Science Department, Carnegie Mellon University, Pittsburgh, PA, USA \and \url{https://www.harrisongrodin.com}}{hgrodin@cs.cmu.edu}{https://orcid.org/0000-0002-0947-3520}{}
\author{Robert Harper}{Computer Science Department, Carnegie Mellon University, Pittsburgh, PA, USA \and \url{https://www.cs.cmu.edu/~rwh/}}{rwh@cs.cmu.edu}{https://orcid.org/0000-0002-9400-2941}{}

\authorrunning{H. Grodin and R. Harper}
\Copyright{Harrison Grodin and Robert Harper}



\supplement{~}
\supplementdetails[cite={niu-sterling-grodin-harper>calf},swhid={swh:1:dir:7750187b111d75acca1980e9abffae2d63ffbe69;origin=https://github.com/jonsterling/agda-calf;visit=swh:1:snp:970335e4251b980e4c493d29d8bb274821ef4d1e;anchor=swh:1:rev:75627cd7a08bc41ab32820eba6e3cc2d4573211a},subcategory={Source Code}, swhdelimiter={,}]{Software}{https://github.com/jonsterling/agda-calf}

\acknowledgements{We are grateful to Yue Niu, Max New, and David Spivak for insightful discussions about this research.}

\funding{This material is based upon work supported by the United States Air Force Office of Scientific Research under grant number FA9550-21-0009 (Tristan Nguyen, program manager) and the National Science Foundation under grant number CCF-1901381. Any opinions, findings and conclusions or recommendations expressed in this material are those of the authors and do not necessarily reflect the views of the AFOSR or NSF.}

\nolinenumbers

\newcommand{\calf}{\textbf{calf}}
\newcommand{\Set}{\mathbf{Set}}
\usepackage{colonequals}
\usepackage{bbm}
\usepackage{relsize}
\usepackage{mathpartir}
\usepackage{mathtools}

\NewDocumentCommand{\kw}{m}{\mathsf{#1}}
\NewDocumentCommand{\lab}{m}{\mathsf{#1}}
\newcommand{\code}[1]{\lstinline{#1}}

\definecolor{OrangeRed}{RGB}{237,19,90}
\definecolor{Cerulean}{RGB}{0,162,227}
\AtBeginDocument{\colorlet{defaultcolor}{.}}
\NewDocumentCommand{\POSITIVE}{m}{{\color{OrangeRed}{#1}}}
\NewDocumentCommand{\NEGATIVE}{m}{{\color{Cerulean}{#1}}}
\NewDocumentCommand{\DEFAULT}{m}{{\color{defaultcolor}{#1}}}

\DeclarePairedDelimiter{\tuple}{\langle}{\rangle}
\newcommand{\mdoubleplus}{\ensuremath{\mathbin{+\mkern-10mu+}}}

\NewDocumentCommand{\T}{s o}{\POSITIVE{\kw{T}}{\IfBooleanT#1{\POSITIVE{(}}}{\IfValueT{#2}{#2}}{\IfBooleanT#1{\POSITIVE{)}}}}

\NewDocumentCommand{\tyU}{s m}{\POSITIVE{\kw{U}\IfBooleanT#1{\POSITIVE{(}}\NEGATIVE{#2}\IfBooleanT#1{\POSITIVE{)}}}}
\NewDocumentCommand{\tyUnitP}{}{\POSITIVE{1}}
\NewDocumentCommand{\tyProdP}{m m}{\POSITIVE{{#1} \times {#2}}}

\NewDocumentCommand{\tySum}{m m}{\POSITIVE{{#1} + {#2}}}
\NewDocumentCommand{\tyInd}{m m}{\POSITIVE{\mu{#1}.\ {#2}}}

\NewDocumentCommand{\tyConP}{m m}{\POSITIVE{\kw{#1}({#2})}}

\NewDocumentCommand{\tyF}{s m}{\NEGATIVE{\kw{F}\IfBooleanT#1{\NEGATIVE{(}}\POSITIVE{#2}\IfBooleanT#1{\NEGATIVE{)}}}}

\NewDocumentCommand{\tyProdN}{m m}{\NEGATIVE{{#1} \times {#2}}}
\NewDocumentCommand{\tyDSumN}{m m m}{\NEGATIVE{\Sigma}_{\DEFAULT{#2 : {}}\POSITIVE{#1}}{\NEGATIVE{#3}}}
\NewDocumentCommand{\tyCopower}{m m}{\NEGATIVE{\POSITIVE{#1} \ltimes {#2}}}
\NewDocumentCommand{\tyArr}{m m}{\NEGATIVE{\POSITIVE{#1} \to {#2}}}
\NewDocumentCommand{\tyCoi}{m m}{\NEGATIVE{\nu{#1}.\ {#2}}}
\NewDocumentCommand{\CofreeComonad}{o}{\NEGATIVE{D}\IfValueT{#1}{_{\NEGATIVE{#1}}}}
\NewDocumentCommand{\tyConN}{m m}{\NEGATIVE{\kw{#1}({#2})}}
\NewDocumentCommand{\tyQueue}{m}{\tyConN{queue}{#1}}

\NewDocumentCommand{\expProdN}{m m}{\tuple{{#1}, {#2}}}

\NewDocumentCommand{\expCopower}{m m}{\tuple{{#1}, {#2}}}
\NewDocumentCommand{\expArr}{m m}{\lambda {#1}.\ {#2}}
\NewDocumentCommand{\expCoi}{m m m}{\kw{gen}({#1}.\ {#2};\ {#3})}

\NewDocumentCommand{\Step}{o m m}{\kw{step}\IfValueT{#1}{_{\NEGATIVE{#1}}}^{#2}({#3})}

\lstdefinelanguage{agda}{
morekeywords={tp,
F, U,
ret, bind, step,
record, coinductive, field
}
}

\EventEditors{Paolo Baldan and Valeria de Paiva}
\EventNoEds{2}
\EventLongTitle{10th Conference on Algebra and Coalgebra in Computer Science (CALCO 2023)}
\EventShortTitle{CALCO 2023}
\EventAcronym{CALCO}
\EventYear{2023}
\EventDate{June 19--21, 2023}
\EventLocation{Indiana University Bloomington, IN, USA}
\EventLogo{}
\SeriesVolume{270}
\ArticleNo{23}

\begin{document}

\maketitle

\vspace{0.5\baselineskip}
\enlargethispage{-0.5\baselineskip}

\begin{abstract}
Amortized analysis is a program cost analysis technique for data structures in which the cost of operations is specified in aggregate, under the assumption of continued sequential use.
Typically, amortized analyses are presented inductively, in terms of finite sequences of operations.
We give an alternative coinductive formulation and prove that it is equivalent to the standard inductive definition.
We describe a classic amortized data structure, the batched queue, and outline a coinductive proof of its amortized efficiency in \calf{}, a dependent type theory for cost analysis.
\end{abstract}

\section{Introduction}

The \calf{} framework is a dependent type theory that supports verification of both correctness conditions and cost bounds~\cite{niu-sterling-grodin-harper>2022}, based on call-by-push-value~\cite{levy>2003}.
Amortized analysis is a cost analysis technique for data structures in which the operation costs are specified in aggregate, under the assumption of continued sequential use~\cite{tarjan>1985}.
In this work, we demonstrate how amortized analysis can be understood as coalgebraic in \calf{}.

In call-by-push-value, there are two sorts of types: value types $\POSITIVE{A}, \POSITIVE{B}, \POSITIVE{C}$ and computation types $\NEGATIVE{X}, \NEGATIVE{Y}, \NEGATIVE{Z}$.
The type $\tyF{A}$ is a computation type classifying computations that result in a value of type $\POSITIVE{A}$, and the type $\tyU{X}$ is a value type classifying suspended computations of type $\NEGATIVE{X}$.
Computation types beyond $\tyF{A}$ will be essential for amortized analysis; in particular, we will make extensive use of products $\tyProdN{X}{Y}$, coproducts $\tyDSumN{A}{a}{X(\DEFAULT{a})}$, powers $\tyArr{A}{X}$, and coinductive types $\tyCoi{X}{Y(X)}$ \cite{balan-kurz>2010}, all of which are computation types.

Semantically, we will interpret value types in $\Set$ and computation types in the category of $\mathbb{C}$-sets, where $\mathbb{C}$ is a monoid representing cost, as is standard for cost analysis of functional programs~\cite{danielsson>2008,danner-licata-ramyaa>2015,kavvos-morehouse-licata-danner>2019,cutler-licata-danner>2020}.
This is a simplification of \calf{}, avoiding modalities.
As in \calf{}, we provide a primitive effect $\Step{c}{-}$ that incurs $c$ units of abstract cost, interpreted using the $\mathbb{C}$-action.
The $\mathbb{C}$-action associated to a computation type justifies equations describing how steps are incorporated into its elements:
\begin{align*}
  \Step[\tyProdN{X}{Y}]{c}{\expProdN{x}{y}} &= \expProdN{\Step[X]{c}{x}}{\Step[Y]{c}{y}} \\
  \Step[\tyDSumN{A}{a}{X(\DEFAULT{a})}]{c}{\expCopower{a}{x}} &= \expCopower{a}{\Step[X]{c}{x}} \\
  \Step[\tyArr{A}{X}]{c}{\expArr{a}{x}} &= \expArr{a}{\Step[X]{c}{x}} \\
  \Step[\tyCoi{X}{Y(X)}]{c}{\expCoi{a}{y}{a_0}} &= \expCoi{a}{\Step[Y(\tyCoi{X}{Y(X)})]{c}{y}}{a_0}
\end{align*}
In other words, cost at a product or power type is incurred pointwise, cost at a coproduct type is pushed into the given summand, and cost at a coinductive type is propagated forward.
In this work, we will make use of the $\POSITIVE{A}$-wide coproduct of a computation type $\NEGATIVE{X}$, also known as the \emph{copower} of $\NEGATIVE{X}$ by $\POSITIVE{A}$~\cite{kelly>1982,egger-mogelberg-simpson>2009}, which we write as $\tyCopower{A}{X} \triangleq \tyDSumN{A}{-}{\NEGATIVE{X}}$.
Note that $\tyCopower{\tyUnitP}{X}$ is isomorphic to $\NEGATIVE{X}$.

\section{Cofree Comonads for Amortized Abstract Data Types}\label{sec:cofree-comonad-adt}

Throughout this paper, we will use queues as a running example of an abstract data type, although the development generalizes to other sequential-use abstract data types.
Queues are an abstract type representing an ordered collection with a first-in-first-out data policy.
Let value type $\POSITIVE{E}$ be the type of elements; the queue operations can be written as follows:
\begin{align*}
  \lab{enqueue}[e] &\sim \tyUnitP{} \\
  \lab{dequeue} &\sim \tySum{E}{\tyUnitP{}}
\end{align*}
This signature describes an operation $\lab{enqueue}[e]$ for each $e : \POSITIVE{E}$ and an operation $\lab{dequeue}$.

In a type theory with one sort of type, a machine offering these operations is given via the following cofree comonad~\cite{jacobs>1995,power-shkaravska>2004,plotkin-power>2008}, interpreted in $\Set{}$:
{
\renewcommand{\POSITIVE}[1]{#1}
\renewcommand{\NEGATIVE}[1]{#1}
\[ \tyQueue{X} \triangleq \tyCoi{Q}{\tyProdN{(\lab{quit} : \NEGATIVE{X})}{\tyProdN{(\lab{enqueue} : \tyArr{E}{Q})}{(\lab{dequeue} : \tyProdP{(\tySum{E}{\tyUnitP})}{Q})}}} \]
}
Up to isomorphism, each operation corresponds to a product of its output type and $Q$, using a function for an $E$-wide product.
In call-by-push-value, though, we must distinguish between a product of computation types and a copower of a value type and a computation type.
Since the result type of an operation is a value type, such as $\tySum{E}{\tyUnitP{}}$ for the $\lab{dequeue}$ operation, we must use the latter.
Thus, we may define the type of (amortized) queues as follows, interpreted in the category of $\mathbb{C}$-sets:
\[ \tyQueue{X} \triangleq \tyCoi{Q}{\tyProdN{(\lab{quit} : \NEGATIVE{X})}{\tyProdN{(\lab{enqueue} : \tyArr{E}{Q})}{(\lab{dequeue} : \tyCopower{(\tySum{E}{\tyUnitP})}{Q})}}} \]
The type $\tyQueue{X}$ can be understood as ``object-oriented''~\cite{cook>1991,jacobs>1996-objects-coalgebraically,cook>2009}, since the use of a queue involves a sequence of $\lab{enqueue}$ and $\lab{dequeue}$ projections terminated by a $\lab{quit}$.
Cost incurred at this type is propagated forward, accumulating at all future $\lab{quit}$ components (of type $\NEGATIVE{X}$) for end-of-use accounting.

\section{Coinductive Amortized Analysis}

Let $\mathbb{C} = (\mathbb{N}, +, 0)$.
We define two queue implementations of type $\tyQueue{X}$ and prove their amortized equivalence.
Here, we let $\NEGATIVE{X} = \tyF{\tyUnitP}$, requiring that the queues terminate with an element of $\tyF{\tyUnitP}$ (i.e., simply a cost in $\mathbb{C}$).

\begin{example}[Specification Queue]
\begin{lstlisting}[float,caption={Single-list specification implementation of a queue.},label=code:list-queue]
spec-queue : list E → queue (F unit)
quit    (spec-queue l) = ret triv
enqueue (spec-queue l) e = step 1 (spec-queue (l ++ [ e ]))
dequeue (spec-queue []) = ret (nothing , spec-queue [])
dequeue (spec-queue (e ∷ l)) = ret (just e , spec-queue l)
\end{lstlisting}

One simple implementation of a queue, called \code{spec-queue}, is given in \cref{code:list-queue} by coinduction using copattern matching~\cite{abel-pientka-thibodeau-setzer>2013}, using a single list as the underlying representation type.
The enqueue operation is annotated with one unit of cost; however, this is unrealistic, since a full traversal of the list is performed for each enqueue operation.
We will treat this implementation as a client-facing specification, next defining a queue that actually implements this cost model.
\lipicsEnd
\end{example}

\begin{example}[Batched Queue]
\begin{lstlisting}[float,caption={Amortized-efficient batched implementation of a queue.},label=code:batched-queue]
batched-queue : list E → list E → queue (F unit)
quit    (batched-queue bl fl) = step (Φ (bl , fl)) (ret triv)
enqueue (batched-queue bl fl) e = batched-queue (e ∷ bl) fl
dequeue (batched-queue bl []) with reverse bl
... | [] = ret (nothing , batched-queue [] [])
... | e ∷ fl = step (length bl) (ret (just e , batched-queue [] fl))
dequeue (batched-queue bl (e ∷ fl)) =
  ret (just e , batched-queue bl fl)
\end{lstlisting}
Now, we define an amortized-efficient implementation which only incurs one large cost infrequently~\cite{gries>1989,hood-melville>1981,burton>1982,okasaki>thesis}.
This underlying representation type of the implementation is two lists: the ``front list'', $\mathsf{fl}$, and the ``back list'', $\mathsf{bl}$.
Elements are enqueued to $\mathsf{bl}$ and dequeued from $\mathsf{fl}$; if $\mathsf{fl}$ is empty when attempting to dequeue, the current $\mathsf{bl}$ is reversed and used in place of $\mathsf{fl}$ going forward.
The \calf{} implementation, called \code{batched-queue}, is shown in \cref{code:batched-queue}.
The $\lab{quit}$ case uses a \emph{potential function} $\Phi(\mathsf{bl}, \mathsf{fl}) = \mathsf{length}(\mathsf{bl})$, as in the physicist's method of amortized analysis~\cite{tarjan>1985}, accounting for elements enqueued on $\mathsf{bl}$ that were never moved to $\mathsf{fl}$.
\lipicsEnd
\end{example}

\enlargethispage{-\baselineskip}
The amortized analysis is proved via a bisimulation; the theorem statement is analogous to the traditional amortized analysis, using the potential function to accumulate payment~\cite{tarjan>1985}.
Every $\lab{enqueue}$ to \code{spec-queue} pushes one unit of cost forward, while \code{batched-queue} pushes $\mathsf{length}(\mathsf{bl})$ units of cost forward only on the occasional $\lab{dequeue}$, retroactively using its surplus potential from previous $\lab{enqueue}$ operations.

\begin{theorem}[Amortized Analysis of Batched Queue]
  For all lists $\mathsf{bl}$ and $\mathsf{fl}$,
  \[ \text{\normalfont\code{batched-queue}}\ \mathsf{bl}\ \mathsf{fl} = \Step{\Phi (\mathsf{bl} , \mathsf{fl})}{\text{\normalfont\code{spec-queue}}\ (\mathsf{fl} \mdoubleplus \mathsf{reverse}\ \mathsf{bl})}. \]
\end{theorem}
\begin{proof}
  By routine coinduction, propagating cost forward over computation types.
\end{proof}

\section{Relation to Inductive Amortized Analysis}

Amortized analysis is typically framed algebraically, describing the cost incurred by a finite sequence of operations.
In the preceding sections we observed that the analysis is naturally viewed as \emph{coalgebraic}.
In fact these perspectives are equivalent.
Define the free monad corresponding to the queue operation signature given in \cref{sec:cofree-comonad-adt}:
\[ \tyConP{program}{A} \triangleq \tyInd{P}{\tySum{(\lab{return} : A)}{\tySum{(\lab{enqueue} : \tyProdP{E}{P})}{(\lab{dequeue} : \tyU*{\tyArr{\tySum{E}{\tyUnitP}}{\tyF{P}}})}}} \]
An element of $\tyConP{program}{A}$ is a finite sequence of queue instructions terminated by returning a value of type $\POSITIVE{A}$.
We may evaluate a program on a queue, by induction on the program:
\[ \kw{eval} : \tyArr{\tyConP{program}{A}}{\tyArr{\tyU*{\tyQueue{X}}}{\tyCopower{A}{X}}} \]
\begin{lstlisting}[float,caption={Program evaluation at a queue.},label=code:evaluate-program]
eval : queue-program A → U (queue X) → A ⋉ X
eval (return a   ) q = a , Queue.quit q
eval (enqueue e p) q = eval p (Queue.enqueue q e)
eval (dequeue k  ) q =
  bind (k (proj₁ (Queue.dequeue q))) λ p →
  eval p (proj₂ (Queue.dequeue q))
\end{lstlisting}
This expresses the usual notion of running a sequence of operations on a data structure; the code is in \cref{code:evaluate-program}.
Semantically, this definition corresponds to a morphism \[ \tyCopower{\tyConP{program}{A}}{\tyQueue{X}} \to \tyCopower{A}{X} \] resembling a monad-comonad interaction law~\cite{plotkin-power>2008,katsumata-rivas-uustalu>2020}, here adjusted for call-by-push-value.
Using $\kw{eval}$, we may define an alternative notion of queue equivalence.
Let $q_1, q_2 : \tyConN{queue}{X}$:
\begin{definition}[Sequence-of-Operations Queue Equivalence]
  Say $q_1 \approx q_2$ iff for all types $\POSITIVE{A}$ and programs $p : \tyConP{program}{A}$, it is the case that $\kw{eval} (p, q_1) = \kw{eval} (p, q_2)$.
\end{definition}
\begin{theorem}[Amortizing Sequences of Operations]
  It is the case that $q_1 = q_2$ iff $q_1 \approx q_2$.
\end{theorem}
\begin{proof}
  By routine $(\Rightarrow)$ induction and $(\Leftarrow)$ coinduction.
\end{proof}
Thus, coalgebraic amortized equivalence coincides with the traditional algebraic notion.
Unsurprisingly, a proof that $q_1 \approx q_2$ shares the same core reasoning as a proof that $q_1 = q_2$; however, it requires the auxiliary definitions of $\tyConP{program}{A}$ and $\kw{eval}$.

\section{Conclusion}

Here, we developed a computation type of amortized queues in \calf{} as the cofree comonad of a functor based on the product, power, and copower computation type constructors, built to propagate cost forward for end-of-use accounting.
We defined specification and amortized queue implementations and stated a theorem relating them via the physicist's method of amortized analysis.
Finally, we observed that coinductive bisimulation coincides with traditional sequence-of-operations reasoning in amortized analysis.
Our results for queues and two other simple amortized data structures are formalized in \calf{}, which is embedded in Agda~\cite{niu-sterling-grodin-harper>calf}.

In future work, we hope to extend this approach to support abstract data types with binary and parallel operations, infinite sequences of operations, and situations in which an amortized implementation may be less costly than the specification.
Additionally, we hope to better characterize the given constructions, accounting for the asymmetry present in call-by-push-value.
As abstract data types are described via a comonad on the category of algebras for a monad, we also hope to connect to bialgebraic presentations of operational semantics~\cite{turi-plotkin>1997}.

%%
%% Bibliography
%%

%% Please use bibtex,

\enlargethispage{\baselineskip}
\bibliography{amortized}

\end{document}